\documentclass[showpacs,aps,prl,twocolumn,superscriptaddress]{revtex4-1}

\usepackage{graphicx}
\usepackage{hyperref}
\usepackage{epstopdf}
\usepackage{dcolumn}
\usepackage{bm}
\usepackage{amsmath}
\usepackage{amssymb}
\usepackage[dvipsnames]{xcolor}
\usepackage{color} 
\usepackage{ulem}
\usepackage{xspace}

\renewcommand{\vec}[1]{\boldsymbol{#1}}

\newcommand{\las}[0]{\langle}
\newcommand{\ras}[0]{\rangle}

\newcommand{\oh}{\mbox{$\frac{1}{2}$}}

\newcommand{\ve}[1]{\boldsymbol{#1}}

\setcitestyle{square,numbers}
\usepackage{here}

\begin{document}

\title{Topological terms on topological defects: a quantum Monte Carlo study}

\author{Toshihiro Sato}
\affiliation{\mbox{Institut f\"ur Theoretische Physik und Astrophysik, Universit\"at W\"urzburg, 97074 W\"urzburg, Germany}}
\author{Martin Hohenadler}
\affiliation{\mbox{Institut f\"ur Theoretische Physik und Astrophysik, Universit\"at W\"urzburg, 97074 W\"urzburg, Germany}}
\author{Tarun Grover}
\affiliation{\mbox{Department of Physics, University of California at San Diego, La Jolla, CA 92093, USA}}
\author{John McGreevy}
\affiliation{\mbox{Department of Physics, University of California at San Diego, La Jolla, CA 92093, USA}}
\author{Fakher F. Assaad}
\affiliation{\mbox{Institut f\"ur Theoretische Physik und Astrophysik, Universit\"at W\"urzburg, 97074 W\"urzburg, Germany}}
\affiliation{\mbox{W\"urzburg-Dresden Cluster of Excellence ct.qmat, Am Hubland, 97074 W\"urzburg, Germany}}

\begin{abstract}
Dirac fermions in $2+1$ dimensions with dynamically generated anticommuting
SO(3) antiferromagnetic (AFM) and Z$_2$ Kekul\'e valence-bond solid (KVBS)
masses  map onto   a field theory with a topological $\theta$-term.  This
term provides a mechanism for  continuous phase transitions between different
symmetry-broken states: topological defects of one phase carry the charge of
the other and proliferate at the transition. The $\theta$-term implies that a
domain wall of the Z$_2$ KVBS order parameter harbors a spin-$1/2$ Heisenberg
chain, as described by a $1+1$ dimensional SO(3) non-linear sigma model with
$\theta$-term at $\theta = \pi$. Using pinning fields to stabilize the domain
wall, we show that our auxiliary-field quantum Monte Carlo simulations indeed
support the emergence of a spin-$1/2$ chain at the Z$_2$ topological defect.
This concept can be generalized to higher  dimensions  where   $2+1$
dimensional SO(4) or SO(5) theories   with topological terms  are  realized
at a domain wall.
\end{abstract}

\maketitle

{\it Introduction.}---Topological terms in field theories  play an important role in our
understanding of phases and critical phenomena.      For instance, the
differences  between integer and half-integer spin-$S$ chains  are  a
consequence of   the    $ 2 \pi i  S $  pre-factor of the  integer-valued $\theta$-term  that counts the winding of  a unit vector  over the
sphere.   Dirac fermions provide a  very appealing route to define models
that map onto field theories with topological terms
\cite{Abanov00,Tanaka05,Senthil06,Lee_Sachdev15,Grover08,SatoT17,Li17,Ippoliti18-1,Liu19-1}.
Consider  8-flavored Dirac fermions in $2+1$ dimensions akin to  graphene.  In
this case, there is a maximum of five anti-commuting mass terms   that could,
for instance, correspond to an antiferromagnet (AFM) with three mass terms
and a Kekul\'e valence bond solid (KVBS) with two mass terms \cite{Ryu09}.  The ten commutators of these mass terms correspond to the
generators of the SO(5) group so that Dirac fermions  Yukawa-coupled to
these five mass terms  possess an SO(5) symmetry.      In the
massive phase, one can integrate out the fermions to obtain a
Wess-Zumino-Witten (WZW)  topological term  \cite{Abanov00,Tanaka05} that is
believed to be at the origin of  deconfined quantum criticality
(DQC)~\cite{Senthil04_1,Senthil04_2}.   In particular, it formalizes the
Levin-Senthil picture~\cite{Levin04} of a vortex of the Kekul\'e order
harboring an emergent spin-$1/2$  degree of freedom.

The aim of this Letter is  to demonstrate numerically the consequences of
topological terms in the corresponding field theory. We will do so by
considering a model of Dirac fermions in 2+1 dimensions   with reduced
spatial symmetries such that  the three AFM  and one of the two KVBS mass
terms are dynamically generated. Contrary to the generic KVBS state with a
spontaneously broken U(1)  symmetry in the continuum, our  KVBS state
spontaneously breaks a  Z$_2$ symmetry.   We will  refer to this state as Z$_2$
KVBS.   Starting from the  WZW topological term,  this symmetry reduction
amounts to setting one component of the  five-dimensional field to zero. This
maps the  WZW  term to a $\theta$-term at $\theta=\pi$~\cite{Senthil06}.
Let us assume that the phase transition observed numerically  between the AFM
and the  Z$_2$ KVBS is continuous and  captured by the aforementioned field
theory.  Then,  the  $\theta$-term  leads to the prediction that  in the
Z$_2$ KVBS phase \textit{close} to the transition, a  Z$_2$ KVBS  domain
wall  harbors a spin-$1/2$ chain.     In what follows, we will provide a model---amenable to large scale negative-sign-free auxiliary-field quantum Monte
Carlo (QMC) calculations---that provides compelling  results supporting this
field-theory picture.

\begin{figure}
\centering
\centerline{\includegraphics[width=0.45\textwidth]{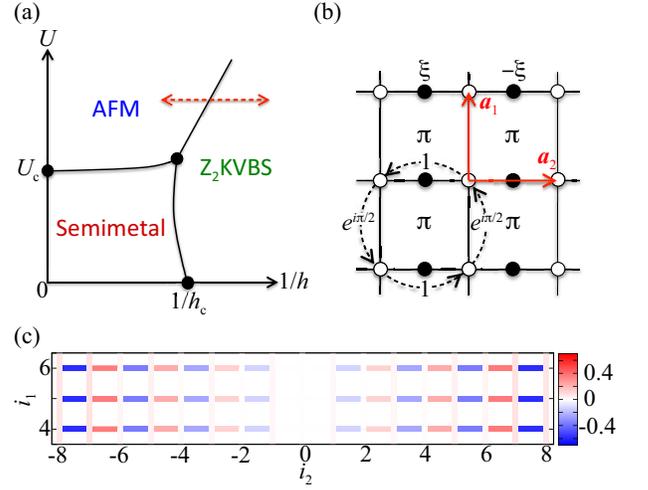}
}
\caption{\label{fig:model}
(a) Schematic ground-state phase diagram with semimetallic, AFM, and Z$_2$ KVBS phases~\cite{SatoT17}. Results correspond to scans along the dashed line.
(b) In our model, fermions  acquire a  $\pi$-flux  when circulating around a plaquette.  Bonds  labeled by a solid circle accommodate an Ising spin that  couples 
to the fermions with magnitude  $\pm\xi$.  
We consider periodic  (open) boundary conditions in the $\vec{a}_1$ ($\vec{a}_2$) directions and freeze the Ising spins on the open boundary to impose a domain wall.
(c) Real-space bond energy change $\Delta \hat{B}_{\vec{i},\vec{a}_l}$ (see text). Here, $L_1=30$ and $L_2=17$.
}
\end{figure}

\textit{ Field theory.}---A theory that accounts for the phase diagram presented in Ref.~\cite{SatoT17} (see Fig.~\ref{fig:model}(a)) contains Dirac fermions Yukawa-coupled
to the   AFM   and Z$_2$ KVBS  mass terms, as described by
\begin{eqnarray}
    {\cal L}_\text{F}   & & = {\Psi}^{\dagger}\left[
     \partial_{\mu} ( \boldsymbol{1}_2 \otimes\gamma_{0}\gamma_{\mu}) +
  \begin{pmatrix}\eta_{\alpha} \\ \chi \end{pmatrix}\cdot
  \begin{pmatrix}{\sigma}_{\alpha} \otimes\gamma_{0} \\ \boldsymbol{1}_2 \otimes i \gamma_{0}\gamma_{5}
   \end{pmatrix}
   \right] {\Psi}^{\phantom\dagger}. \nonumber\\ 
  \label{Field_theory.eq}
\end{eqnarray}
Here, the Dirac spinors ${\Psi}^{\dagger}$ carry a sublattice index, a spin
index, and a valley index.
The $\gamma$-matrices act on the valley and sublattice spaces and satisfy the Clifford algebra $\{\gamma_a,\gamma_b\}=2 \delta_{ab}$.
${\sigma}_{\alpha}$ with $\alpha=1,2,3$ denote the Pauli spin-$1/2$ matrices.
The fact that the SO(3) AFM mass terms ${\sigma}_{\alpha} \otimes\gamma_{0}$
and Z$_2$ KVBS mass terms $\boldsymbol{1}_2 \otimes i \gamma_{0}\gamma_{5}$
anti-commute results in an SO(4)  invariance of the fermionic action: a
global SO(4) rotation of the four-component field
$\vec{\phi}=(\phi_{1},\phi_{2},\phi_{3},\phi_{4})=(\vec{\eta},\chi)$   is
equivalent to a canonical transformation of the fermion operators.       The
dynamics of the field is  governed by   a  four-component  $\varphi^4$ action,
${\cal L}_\text{B}$.   While  ${\cal L}_\text{F}$ has SO(4) symmetry, ${\cal L}_\text{B}$
inherits the SO(3) $ \times $ Z$_2$ symmetry of the  lattice model.

The Lagrangian ${\cal L}={\cal L}_\text{F}+{\cal L}_\text{B}$ can account for many phase transitions.  The Gross-Neveu transitions from semimetal to AFM  or from semimetal to 
Z$_2$ KVBS involve a closing of the mass gap    corresponding to the norm of
the field $\vec{\phi}$.    On the other hand,  QMC simulations (see
Ref.~\cite{SatoT17} and the Supplemental Material (SM)) point to a continuous
transition between the AFM and Z$_2$ KVBS states with an emergent SO(4)
symmetry.  Importantly,  the numerical results  show that the single-particle
gap remains finite across the transition. In the field theory, this implies
that amplitude fluctuations of  $\vec{\phi}$   are frozen and only phase
fluctuations of the field need to be retained. Since the fermions
remain massive, they can be integrated out (in the large mass limit) to obtain
\begin{eqnarray}
S =  \int dx^2 d \tau \frac{1}{g}(\partial_u \hat{\vec{\phi}})^2+i\theta{\cal Q} ~~,~~\theta=\pi
  \label{action-1}
\end{eqnarray}
with
\begin{eqnarray}
{\cal Q}=  \frac{1}{12\pi^2}\int dx^2 d \tau \epsilon_{i,j,k}\epsilon_{\alpha,\beta,\gamma,\delta}\hat{{\phi}}_{\alpha}\partial_i \hat{{\phi}}_{\beta} \partial_j \hat{{\phi}}_{\gamma} \partial_k \hat{{\phi}}_{\delta}.
  \label{Q-1}
\end{eqnarray}
Here, $\hat{\vec{\phi}}(\vec{x},\tau)=\vec{\phi}/|\vec{\phi}| $ defines a
mapping from $2+1$ dimensional Euclidean space-time to the three-dimensional sphere $S^3$.  For smooth field configurations with no 
singularities at infinity ${\cal Q}$ is quantized to  integer values and corresponds to the winding of the unit four-vector $\hat{\vec{\phi}}$ on the hypersphere in four dimensions, $S^3$.

We now consider a smooth domain wall of the Z$_2$ KVBS order parameter.  Such a configuration  can  be obtained by pinning the field   $\hat{\ve{\phi}}$  at the origin and at infinity: $\hat{\vec{\phi}}(\vec{0})=(0,0,0,1)$ and $\hat{\vec{\phi}}(\vec{\infty})=(0,0,0,-1)$.   Specifically,  let us  parameterize   2+1 dimensional  Euclidean space-time with spherical coordinates,   $(\vec{x},\tau) =  r \vec{n}$  with $\vec{n}$ a unit vector, and choose 
\begin{equation}
	\hat{\phi}(r \vec{n})   =  \left(  \sin(f(r)) \vec{n}, \cos(f(r) \right).
\end{equation}
Here,   $f$ is a one-to-one  smooth  function with boundary conditions  $f(0) = 0$ and $f(\infty) = \pi$ and  describes the profile of the domain wall. 
As  shown in  the SM,  the integration over $r$  can now be carried out  to obtain the domain-wall action:
\begin{eqnarray}
S_{DW} =  \int dx d \tau \frac{1}{g}(\partial_u \vec{n})^2+i\pi{\cal Q}_{DW}(\vec{n})
  \label{action-2}
\end{eqnarray}
with
\begin{eqnarray}
{\cal Q}_{DW}(\vec{n}) =  \frac{1}{4\pi}\int dx d\tau \vec{n}\cdot \partial_\tau \vec{n}\times\partial_x\vec{n}.
  \label{W-1}
\end{eqnarray}
Above we have mapped  $S^2$   (on which  $\vec{n}$ is defined)  to $\mathbb{R}^2$.
While the  topological term is independent of the choice of the profile of
the domain wall, $g$ depends on $f$.
The action  in Eq.~(\ref{action-2})   corresponds to that of the spin-$1/2$ Heisenberg chain~\cite{Haldane83,mudry_book}.  
Thereby, the topological $\theta$-term at $\theta=\pi$ has the important consequence that a domain wall of the Z$_2$ KVBS order parameter harbors a spin-$1/2$ Heisenberg chain.    

\textit{ Model.}---The QMC simulations presented in  Ref.~\cite{SatoT17} for
the honeycomb lattice support a direct and continuous transition between the
AFM and Z$_2$ KVBS with an emergent  SO(4)   symmetry and, in principle,
provide a case to test the above predictions.   However,   irrespective of how one places the pinning fields on the honeycomb lattice,  
translation symmetry along the domain wall will be broken.  
Since gaplessness of the spin-$1/2$ chain is protected by a mixed anomaly between 
translations and time-reversal or spin rotations, dimerization along the domain wall will occur. 
Hence, even if  a spin chain   emerges at the domain wall, it will gap out
due to the choice of lattice discretization. To avoid this, we  have
reformulated the model of Ref.~\cite{SatoT17} on   the $\pi$-flux square
lattice.  This provides a lattice discretization of Dirac fermions with a
C$_4$ symmetry (as opposed to C$_3$   for the
honeycomb lattice).  The model Hamiltonian reads
$\hat{H}=\hat{H}_\text{f}+\hat{H}_\text{s}+\hat{H}_\text{fs}$ (see
Fig.~\ref{fig:model}(b))  with
\begin{eqnarray}
  \label{LM}
\hat{H}_\text{f} &  &=\sum_{\langle \vec{ij} \rangle,\sigma}t_{\vec{ij}}\hat{c}_{\vec{i}\sigma}^\dagger \hat{c}^{}_{\vec{j}\sigma}+U\sum_{\vec{i}}(\hat{n}_{\vec{i}\uparrow}-\oh)(\hat{n}_{\vec{i}\downarrow}-\oh),  \\ 
\hat{H}_\text{s}& & =J\sum_{\las\vec{ ij,kl} \rangle}\hat{s}_{\vec{ij}}^{z} \hat{s}_{\vec{kl}}^{z}-h\sum_{\las \vec{ij}\ras}\hat{s}_{\vec{ij}}^{x},  \, \, \,  
\hat{H}_\text{fs} =\sum_{\langle \vec{ ij} \rangle,\sigma}t_{\vec{ij}}\xi_{\vec{ij}} \hat{s}_{\vec{ij}}^{z} \hat{c}_{\vec{i}\sigma}^\dagger \hat{c}^{}_{\vec{j}\sigma}. \nonumber
\end{eqnarray}
While $\hat{H}_\text{f} $ corresponds to the half-filled Hubbard model on the
$\pi$-flux square lattice, $\hat{H}_\text{s}$ is a ferromagnetic,
transverse-field Ising model defined on the bonds $\las \vec{ij} \ras$ of the
square lattice.
$\hat{H}_\text{fs}$ accounts for the coupling between Dirac fermions and Ising spins.
The Hubbard interaction and the fermion-spin coupling can dynamically generate SO(3) AFM order and Z$_2$ KVBS order (ferromagnetic order of the Ising spins), respectively.
For the numerical simulations we used the ALF (Algorithms for Lattice
Fermions) implementation \cite{ALF_v1} of the well-established
finite-temperature auxiliary-field QMC
method~\cite{Blankenbecler81,Assaad08_rev}.
Our model can be simulated without encountering the negative sign problem. 
Henceforth, we use $t=1$ as the energy unit, set $J=-1$, $\xi=0.5$, and $U=7$.
An inverse temperature $\beta=30$ (with Trotter discretization $\Delta\tau=0.1$) yields results representative of the ground state. 
QMC results on torus geometries detailed in the SM suggest  a continuous AFM--Z$_2$ KVBS transition with emergent SO(4) symmetry  at $1/h_c\approx 0.270$.

To pin a domain wall configuration,   we consider a  cylindrical  geometry  and freeze the Ising spins at the edges to  $\hat{s}_{(i_1,-n), (i_1,-n+1)}^{z}=1$ and
 $\hat{s}_{(i_1,n-1), (i_1,n)}^{z}=-1$ where  $L_2 = 2n+1$.    Importantly,
 and taking into account the gauge freedom  to define the $\pi$-flux model,
 translation symmetry by  $\vec{a}_1$ is present.   The model with pinning
 fields has a mirror symmetry corresponding to the combined transformations
 $\hat{c}_{(i_1, i_2),\sigma}^\dagger \to \hat{c}_{(i_1, - i_2),  \sigma}^\dagger$ and $\hat{s}_{(i_1,i_2),(i_1, i_2 +1) }^{z} \to - \hat{s}_{(i_1,i_2),(i_1,-i_2-1)}^{z}$.

\textit{Numerical results.}---To detect the profile of the domain wall,  we measure the bond kinetic energy  $\Delta \hat{B}_{\vec{i},\vec{a}_l}=\langle\hat{B}_{\vec{i},\vec{a}_l}\rangle-\langle\bar{B}\rangle$.
Here, $\hat{B}_{\vec{i},\vec{a}_l}=\sum_\sigma
t_{\vec{i},\vec{i}+\vec{a}_l}(\hat{c}_{\vec{i} \sigma}^\dagger
\hat{c}^{}_{\vec{i}+\vec{a}_l
  \sigma}+\hat{c}_{\vec{i+\vec{a}_l}\sigma}^\dagger
\hat{c}^{}_{\vec{i}\sigma})$ and $\bar{B}=(2L_1L_2)^{-1}\sum_{\vec{i},l}
\hat{B}_{\vec{i},\vec{a}_l}$ where $l=1,2$.
 Figure~\ref{fig:model}(c)  shows  this quantity.  The aforementioned
 translation and mirror symmetries are readily seen.

\begin{figure}
\centering
\centerline{\includegraphics[width=0.45\textwidth]{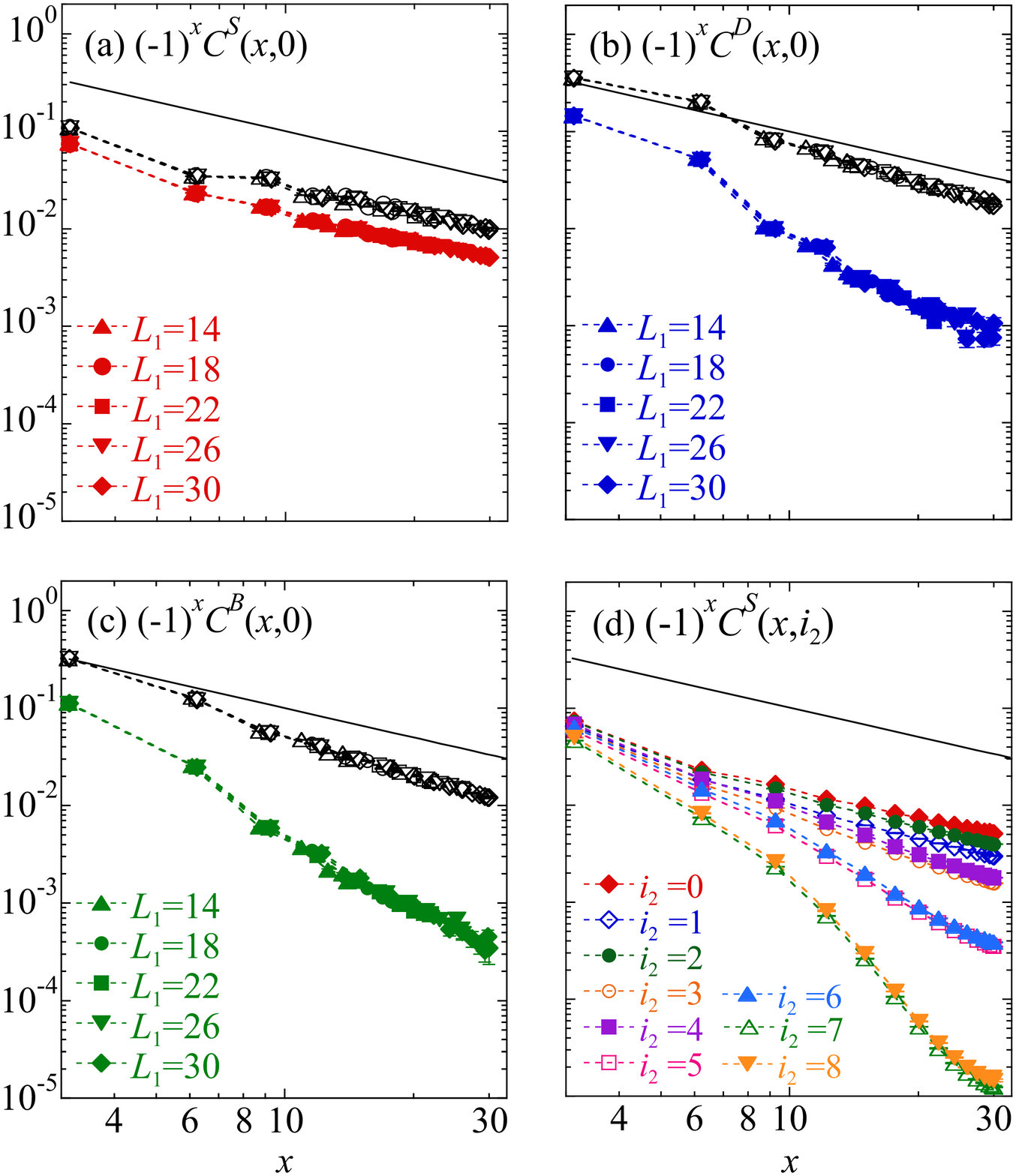}
}
\caption{\label{fig:CF}
Real-space correlation functions of the model~(\ref{LM}) for (a) spin, (b)
dimer, and (c) bond at the domain wall of Z$_2$ KVBS (solid symbols). Here
$L_2=17$ and $x=L_1  {\rm sin}(\pi i_1/L_1)$ is the conformal
distance~\cite{Cardy1996}. Open symbols indicates QMC results for the
one-dimensional Hubbard model with $U/t=4$ and at half filling. (d) Real-space
correlation functions of the model~(\ref{LM}) for spin along the domain
wall. Here, $L_1=30$ and $L_2=17$. Solid line denotes $(-1)^x x^{-1}$.
}
\end{figure}

As discussed above, the field theory of the domain wall is described by an SO(3)  non-linear  sigma model with $\theta$-term at $\theta = \pi$ in 1+1 dimensions. 
We expect this theory to have an emergent SO(4)  \cite{Affleck85,Affleck87}  symmetry  reflecting the fact that spin-spin  and dimer-dimer correlations decay with the same power law  but 
with different logarithmic corrections:    $(-1)^r (\ln{r})^{1/2}r^{-1}$ for the spin ~\cite{Affleck_1989,Singh89,Giamarchi89}   and  $ (-1)^r (\ln{r})^{-3/2}r^{-1}$ for the dimer ~\cite{Giamarchi89}.
In Figs.~\ref{fig:CF}(a-c)   we plot the   
spin  [$C^S(\vec{i})=\langle \hat{\vec{S}}_{\vec{i}}\cdot \hat{\vec{S}}_{\vec{0}}\rangle$], 
dimer [$C^D(\vec{i})=\langle (\hat{\vec{D}}_{\vec{i}}-\langle \hat{\vec{D}}_{\vec{i}} \rangle)\cdot(\hat{\vec{D}}_{\vec{0}}-\langle \hat{\vec{D}}_{\vec{0}} \rangle)\rangle$]  and  
bond [$C^B(\vec{i})=\langle (\hat{B}_{\vec{i},\vec{a}_1}-\langle
\hat{B}_{\vec{i},\vec{a}_1} \rangle)\cdot(\hat{B}_{\vec{0},\vec{a}_1}-\langle
\hat{B}_{\vec{0},\vec{a}_1} \rangle)\rangle$]  correlators as a
function of the conformal distance $x=L_1  {\rm sin}(\pi i_1/L_1)$~\cite{Cardy1996}.  
 Here, $\hat{\vec{S}}_{\vec{i}}=\sum_{\sigma\sigma'} \hat{c}_{\vec{i}
   \sigma}^{\dagger} \vec{\sigma}_{\sigma\sigma'} \hat{c}^{}_{\vec{i}
   \sigma'}$, and $\hat{\vec{D}}_{\vec{i}}=\hat{\vec{S}}_{\vec{i}}\cdot
 \hat{\vec{S}}_{\vec{i}+\vec{a}_1}$.
 The  bond  and dimer correlations  share  the same   symmetries   so that
 we  expect them to decay with the same power law.      We compare our
 results with those for the half-filled Hubbard chain at  $U/t =4$.
 While there is remarkable agreement between the spin correlations  (see
 Fig.~\ref{fig:CF}(a)) it appears that we have to reach  longer length scales
 in the domain wall calculation to observe the $1/r$ power law
 decay for the dimer and bond correlations (see Fig.~\ref{fig:CF}(b),(c)).  A
 possible interpretation  of these numerical  results  is that  the  Hubbard
 model is \textit{closer}  to the emergent SO(4) conformal field theory than the  domain wall
 SO(3)  theory of  Eq.~(\ref{action-2}).
 
The field theory interpretation of the domain wall has consequences.  It
should be independent of the choice of the lattice discretization---provided
that it does not break relevant symmetries such as translation along the
domain wall---and the lattice constant should correspond to a high-energy
scale. In Fig~\ref{fig:CF}(d), we check that the domain wall extends over
many lattice sites in the perpendicular direction.  In particular, for the
value of transverse field $h$ considered, the domain wall extends over
several lattice spacings and the data are consistent with
$C^S(x,i_2)\sim e^{-|i_{2}|/\xi}x^{-1}$ where $\xi \sim 4$.  Varying $1/h$ in
the Z$_2$ KVBS will merely change the profile of the domain wall, thereby
changing the length scale $\xi$ but not the properties of the spin-$1/2$
chain along the domain wall (see SM).  If $1/h$ drops below $1/h_c$ into the
AFM phase, spinons will bind and we expect long-range order to develop within
the domain wall.  Calculations confirming this point of view can be found in
the SM.

\begin{figure}
\centering
\includegraphics[width=0.45\textwidth]{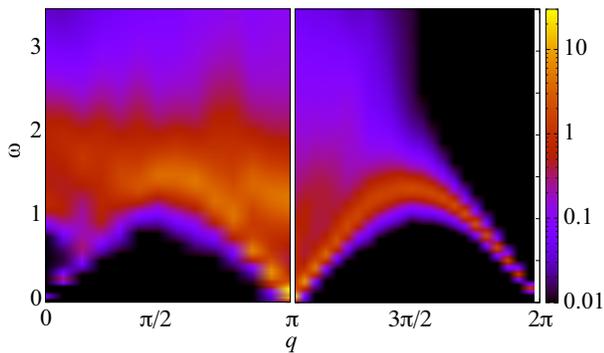} 
\caption{\label{fig:SQ} Dynamical spin structure factor $C^S(q, \omega)$ at
  the domain wall of the Z$_2$ KVBS (left panel). Here, $L_1=26$ and
  $L_2=17$.  From the calculation of the single-particle gap on torus
  geometries (see the SM) we estimate the particle-hole continuum to lie
  above $\omega \simeq 1.5$.  Right panel: dynamical spin structure factor
  for the Hubbard model on a 46-site chain at
  $U/t=4$ and at half filling.}
\end{figure}

Further evidence for an emergent spin-$1/2$ Heisenberg chain can be obtained
from the dynamical spin structure factor at the domain wall of the Z$_2$ KVBS
shown in the left panel of Fig.~\ref{fig:SQ}.  The key feature of the spin-$1/2$
chain is that the low-lying excitations are well described by the two-spinon
continuum revealed by the dynamical spin structure factor.  In the
thermodynamic limit, such excitations have a support in the wave vector, $q$,
versus frequency, $\omega$, plane with lower and upper bounds given by
$\omega_\text{l}(q)\sim\pi/2|\sin(q)|$ and $\omega_\text{u}(q)\sim\pi|\sin(q/2)|$,
respectively.  Experimental as well as theoretical calculations of the
dynamical spin structure factor can be found in Ref.~\cite{Lake13}.
Figure~\ref{fig:SQ} shows the results for the dynamical spin structure factor
$C^S(q, \omega)$ at the domain wall of the Z$_2$ KVBS.  In our QMC
simulations, $C^S(q, \omega)$ was obtained via the analytic continuation of
the imaginary-time-displaced spin correlation functions
$C^S(q, \tau)=\langle
\hat{\vec{S}}_{q}(\tau)\cdot\hat{\vec{S}}_{-q}(0)\rangle={L_1}^{-1}\sum_{i_1}
e^{i q i_1} \langle
\hat{\vec{S}}_{(i_1,0)}(\tau)\cdot\hat{\vec{S}}_{\ve{0}}(0) \rangle $ at
$i_2=0$.  Below the particle-hole continuum, estimated to start at
$\omega > 1.5 t$, our QMC results reproduce the well-known features of the
two-spinon continuum [see the right panel of Fig.~\ref{fig:SQ}].

{\it Summary and discussion.}---We have shown how to probe topological terms
in lattice realizations of field  theories by   pinning defects.   The
explicit example provided in this work is based on a model where the
effective field theory has a   $\theta$-term at $\theta = \pi$ in 2+1
dimensions with emergent SO(4) symmetry.  In the lattice realization of  this
model,  the SO(4) symmetry reduces to SO(3)$\times$ Z$_2$  and  we consider  a
domain wall  of the  Z$_2$ field.  The emergent SO(4) symmetry  then
suggests that the domain wall harbors a spin-$1/2$ chain. Our numerical results
confirm this point of view.

The above argument holds for continuous transitions with emergent symmetries. There is
an ongoing debate on the nature of the generic DQCP between AFM and VBS
\cite{Shao15,Nahum15,Sandvik20} (or quantum spin Hall (QSH) and s-wave
superconducting (SSC) \cite{Liu18}) orders with emergent SO(5) symmetry
\cite{Nahum15_1}.  Compelling evidence for a continuous transition as well as
emergent SO(5) symmetry from finite-size calculations has been put
forward. However the critical exponents stand at odds with the bootstrap
bounds \cite{Poland18}.  To resolve this apparent contradiction, one can conjecture
\cite{WangC17,Nahum19,WangC19} that an SO(5) conformal field theory
indeed exists in spatial dimensions slightly greater than two that, however,
collides with another fixed point and becomes complex upon
tuning the dimension down to two.
Proximity to fixed-point collision is at the origin of a very
slow renormalization group flow and associated very long correlation lengths
\cite{Janssen14,WangC17}.  In fact, recent simulations of the SO(5)
non-linear sigma model with a WZW topological term support this point of view
\cite{WangZ20}.  Very similar arguments can be applied to the present case
where weakly first-order transitions were reported for similar symmetry
classes \cite{Torres19,Serna18,Zhao19}.  Hence, a weakly first-order
transition does not impair the notion that topological terms can play a
dominant role at intermediate length scales.

Our approach can also be applied to other models.  For the AFM-VBS
transition, pinning a C$_4$ vortex on a system with open boundary conditions should
result in a spinon that can be probed via the spin susceptibility.
In the context of the QSH-SSC transition of Ref.~\cite{Liu18} it is
possible to pin a skyrmion of the O(3) QSH order parameter.  Since topology
states that the skyrmion carries charge 2e \cite{Grover08} this should result
in a doping of the system.

Our observation may have some utility in three dimensions, where
there are also pairs of ordered phases for which
the disorder operators for one phase are charged under the
symmetry broken by the other.  A simple example involves a cubic-lattice 
AFM and a cubic-lattice VBS \cite{2010PhRvB..81d5120H}.
As in two dimensions, the skyrmions carry lattice-symmetry quantum numbers,
and the defects of the VBS pattern (which are hedgehogs) carry spin.
This can be encoded in a sigma model with a WZW term, now with softly-broken 
$\mathrm{SO}(6) \supset \mathrm{SO(3)}_\mathrm{AFM} \times \mathrm{SO(3)}_\mathrm{VBS}$ symmetry.
However, in addition to the usual possibility of a first-order transition \cite{2012PhRvB..86m4408B}, 
a direct transition between these two phases can also be 
preempted by an intermediate disordered {\it phase} for the following reason:
in contrast to two dimensions, compact abelian gauge theory with small amounts of 
charged matter (QED) has a (familiar) deconfined phase in three dimensions.  
But, as in the above discussion, the WZW term still has consequences 
{\it within} the ordered phases.  
For definiteness and similarity with our example above, consider breaking the cubic lattice symmetry down to 
$\mathrm{Z}_2 \times \mathrm{Z}_2$, where the second $\mathrm{Z}_2$
 represents reflections in $\hat z$, say.
The associated sigma model then has 
$\mathrm{SO}(5) \supset \mathrm{SO}(3)_\mathrm{AFM}\times (\mathrm{Z}_2 \times \mathrm{Z}_2)_\mathrm{VBS}$ 
symmetry with a $\theta$-term at $\theta=\pi$. In analogy to the case
considered here, the domain wall of the Z$_2$ part of the VBS order parameter 
transverse to the $\hat z$ direction
will host an SO(4) non-linear sigma model at $\theta=\pi$, now in $2+1$ dimensions.
This is a description of the deconfined critical point between AFM and KVBS
orders in two dimensions. Along a domain wall of the VBS pattern, spin and
VBS correlations are predicted to be long-ranged, with the same exponents.

Another possibility 
is to break the cubic lattice symmetry down to $\mathrm{C}_4 \times \mathrm{Z}_2$,
where again the $\mathrm{Z}_2$ represents reflections in $\hat z$.
The associated sigma model then has $\mathrm{SO}(6) \supset \mathrm{SO}(3)_\mathrm{AFM}\times 
(\mathrm{SO}(2) \times \mathrm{Z}_2)_\mathrm{VBS}$ symmetry with a WZW term.
Now, the domain wall of the $\mathrm{Z}_2$ part of the VBS order parameter, transverse to the $\hat z$ direction,
will host an $\mathrm{SO}(5)$ non-linear sigma model with a WZW term, now in $2+1$ dimensions. This construction could provide an alternative 
for the Landau-level projection formulation of this theory \cite{Ippoliti18,WangZ20}.
This is a description of the DQCP between AFM and VBS.
 
\bigskip
\begin{acknowledgments}
We thank J. S. Hofmann, M. Oshikawa,  Z. Wang, C. Xu  and Yi-Zhuang You for insightful discussions.
We thank the Gauss Centre for Supercomputing (SuperMUC at the Leibniz
Supercomputing Centre) for generous allocation of supercomputing resources.
The research has been supported by the Deutsche Forschungsgemeinschaft through grant numbers AS 120/15-1 (TS),  AS120/14-1  (FFA),  the W\"urzburg-Dresden Cluster of  excellence on Complexity and Topology in Quantum Matter - ct.qmat (EXC 2147, project-id 39085490)   (FFA) and SFB 170 ToCoTronics  (MH).  TG is supported by the National Science Foundation under Grant No. DMR-1752417, and as an Alfred P. Sloan Research Fellow. FFA and TG thank the  BaCaTeC for partial financial support.

\end{acknowledgments}

\clearpage
\section{Supplemental Material}
In this supplemental material we will first derive the domain wall action: a
(1+1) dimensional SO(3) non-linear sigma model with $\theta$-term at
$\theta = \pi$. Next we will review our auxiliary-field quantum Monte Carlo
(AFQMC) simulations on the torus. They support the point of view that---on
the accessible lattice sizes--our model indeed shows a continuous
transition between an SO(3) antiferromagnet (AFM) and Z$_2$ KVBS with
emergent SO(4) symmetry.  We then provide more data on the profile of the
domain wall and spin dynamics along the domain wall when tuning through the
transition. Finally, we will show that AFQMC simulations of the
one-dimensional repulsive Hubbard model at half filing reproduce the expected
power-law decay of real-space spin, dimer, and bond correlation functions.

\section{ Spin-1/2  chain  pinned at the Z$_2$  domain wall:  a field theoretic approach}
\label{Field_theory.sec}
In this section, we will detail the calculation that shows that   starting from the   $\theta$-term at $\theta=\pi$,   a domain wall  of $\mathrm{Z}_2$ KVBS nucleates  a spin-$1/2$ chain.  Our starting point is the field  theory  discussed in the main text: 
\begin{eqnarray}
S =  \int dx^2 d \tau \frac{1}{g}(\partial_u \hat{\ve{\phi}})^2+i\theta{\cal Q} ~~,~~\theta=\pi
  \label{action-1}
\end{eqnarray}
with
\begin{eqnarray}
{\cal Q}=  \frac{1}{12\pi^2}\int dx^2 d \tau \epsilon_{i,j,k}\epsilon_{\alpha,\beta,\gamma,\delta}\hat{{\phi}}_{\alpha}\partial_i \hat{{\phi}}_{\beta} \partial_j  \hat{{\phi}}_{\gamma} \partial_k \hat{{\phi}}_{\delta}.
 \end{eqnarray}
 Here, $\hat{\ve{\phi}}=\hat{\ve{\phi}}(\ve{x},\tau)$ defines a mapping
 between the $2+1$ dimensional Euclidean space-time to the unit sphere in
 four dimensions: $S^3$.  For any smooth field configuration, ${\cal Q}$
 counts the winding of the unit four-vector
 $\hat{\ve{\phi}}=\ve{\phi}/|\ve{\phi}|$ on the hypersphere $S^3$ and takes
 an integer value.  The value of $\theta$ is thereby of great importance.
 One can start from the Wess-Zumino-Witten term that appears in a setting
 where the Dirac fermions couple symmetrically to a quintuplet of
 anti-commuting mass terms and set one mass to zero.  A calculation will
 then lead to the aforementioned $\theta$-term at $\theta=\pi$
 \cite{Senthil06}.  Here, our aim is to consider a domain-wall configuration.
 For the calculation to be well defined, we have to make sure that the field
 configuration $\hat{\ve{\phi}}(\ve{x},\tau)$ has no singularities.  To
 achieve this goal, it is useful to include the point at infinity in
 Euclidean space, so that the base space is topologically a three-sphere.
 We can specify coordinates by embedding the three-sphere into $\mathbb{R}^4$
 Euclidean space by
\begin{eqnarray}
	&& x_1  =  \cos(\varphi_1)     \nonumber \\
	& &  x_2  =  \sin(\varphi_1) \cos(\varphi_2)     \nonumber \\
	& &  x_3 =  \sin(\varphi_1) \sin(\varphi_2) \cos(\varphi_3)     \nonumber \\
	& &  x_4 =  \sin(\varphi_1) \sin(\varphi_2) \sin(\varphi_3)     
\end{eqnarray}
where  $\varphi_1 \in [0,\pi] $,   $\varphi_2 \in [0,\pi] $,  and $\varphi_3 \in [0,2\pi] $.   Since the  topological term is independent  of the choice of the metric we obtain with this  parameterization: 
\begin{eqnarray}
{\cal Q}=  \frac{1}{12\pi^2 } & &  \int_{0}^{\pi} d \varphi_1 \int_{0}^{\pi} d \varphi_2  \int_{0}^{2\pi}  d \varphi_3   \nonumber  \\ 
  & & \epsilon_{i,j,k}\epsilon_{\alpha,\beta,\gamma,\delta}{\hat{\phi}}_{\alpha}\partial_i {\hat{\phi}}_{\beta} \partial_j {\hat{\phi}}_{\gamma} \partial_k {\hat{\phi}}_{\delta}
 \end{eqnarray}
with   $  \partial_i = \frac{\partial}{\partial \varphi_i} $. 
With this  compactification  of $\mathbb{R}^3$    we can define a smooth domain-wall configuration as: 
\begin{eqnarray}
\label{Domain_wall.Eq}
	&& {\hat{\phi}}_{4} =  \cos(f(\varphi_1))  \nonumber \\
	&& {\hat{\phi}}_{3} =  \sin(f(\varphi_1)) n_{3}  \nonumber \\	 
	&& {\hat{\phi}}_{2} =  \sin(f(\varphi_1)) n_{2}  \nonumber \\	 
	&& {\hat{\phi}}_{1} =  \sin(f(\varphi_1)) n_{1}   
\end{eqnarray}
Here $\ve{n}(\varphi_2,\varphi_3) = \left( n_1, n_2, n_3 \right) $.  From
$|\hat{\ve{\phi}}| =1 $ follows that $|\ve{n}| =1 $.  $\varphi_2$ and
$\varphi_3$ define a point on the unit sphere in three dimensions, $S^2$, that
is isomorphic to $\mathbb{R}^2$.  Hence, $\ve{n} $ defines a mapping from
$\mathbb{R}^2$ to the unit sphere in $\mathbb{R}^3$.

We can now return to $\hat{\ve{{\phi}}}$.  The function $f(\varphi_1)$
defines a generic domain wall and is required to satisfy the following
properties.  $f(0) = 0$, $f(\pi) = \pi$ and $ f(\varphi_1)$ is a one-to-one
mapping from $\left[ 0,\pi \right]$ to $\left[ 0,\pi \right]$.  At infinity
in $\mathbb{R}^3$, corresponding to $ \varphi_1 =0 $,
$\hat{\ve{{\phi}}} = (0,0,0,1)$ and at the origin, $ \varphi_1 = \pi $,
$\hat{\ve{{\phi}}} = (0,0,0,-1)$.  The pinning of $\hat{\ve{{\phi}}}$ at the
origin and at infinity defines a Z$_2$ domain wall in the fourth component of
the field.  This choice of the field $\hat{\ve{\phi}}$ explicitly breaks
SO(4) symmetry down to SO(3) corresponding to rotations of the $\ve{n}$
vector.

For the domain-wall field configuration of Eq.~\ref{Domain_wall.Eq}, we can
evaluate the $\theta$-term by explicitly carrying out the integration over
$\varphi_1$.  Since $f(\varphi_1) $ is a one-to-one mapping and ${\cal Q} $
is a topological quantity, the value of ${\cal Q} $ for the domain wall field
of Eq.~\ref{Domain_wall.Eq} is identical to the deformed domain wall:
 \begin{eqnarray}
\label{Domain_wall1.Eq}
	&& \hat{\phi}_{4} =  \cos(\varphi_1)  \nonumber \\
	&& \hat{\phi}_{3} =  \sin(\varphi_1) n_{3}  \nonumber \\	 
	&& \hat{\phi}_{2} =  \sin(\varphi_1) n_{2}  \nonumber \\	 
	&& \hat{\phi}_{1} =  \sin(\varphi_1) n_{1}.  
\end{eqnarray}
Inserting this result in  ${\cal Q}  $   and evaluating the  integral over $\varphi_1$ gives 
\begin{equation}
	{\cal Q}_{\text{DW}} = \frac{1}{4 \pi}   \int_{0}^{\pi} d \varphi_2  \int_{0}^{2 \pi }  d  \, \varphi_3 \,    \ve{n} \cdot \partial_{\varphi_2}  \ve{n} \times  \partial_{\varphi_3}  \ve{n}.
\end{equation}
In summary,  the domain-wall action reads
\begin{equation}
S_{DW} =  \int dx d \tau \frac{1}{g}(\partial_u \hat{\ve{n}})^2+i \pi {\cal Q}_{\text{DW}},
\end{equation}
which corresponds to the  coherent spin state path  integral of the spin-$1/2$ chain   \cite{Haldane83,mudry_book}. 

\section{Continuous AFM-Z$_2$ KVBS transition with emergent SO(4) symmetry}
\label{QMC.sec}
As mentioned in the main text,  our calculations in \cite{SatoT17} were carried out on the 
honeycomb lattice.    Here,  we map out the bulk phase diagram  on a torus  geometry for the $\pi$-flux   lattice  and show that the model has a similar phase diagram.

We consider a model of Dirac fermions in $2+1$ dimensions with Hamiltonian $\hat{H}=\hat{H}_\text{f}+\hat{H}_\text{s}+\hat{H}_\text{fs}$ (see Fig.1(b) of the main text):
\begin{eqnarray}
\hat{H}_\text{f} &  &=\sum_{\langle \ve{i}\ve{j} \rangle,\sigma}t_{\ve{i}\ve{j}}\hat{c}_{\ve{i}\sigma}^\dagger \hat{c}^{}_{\ve{j}\sigma}+U\sum_{\ve{i}}
(\hat{n}_{\ve{i}\uparrow}-\oh)(\hat{n}_{\ve{i}\downarrow}-\oh),  \nonumber\\ 
\hat{H}_\text{s}& & =J\sum_{\las \ve{i}\ve{j},\ve{k}\ve{l} \rangle}\hat{s}_{\ve{i}\ve{j}}^{z} \hat{s}_{\ve{k}\ve{l}}^{z}
 -h\sum_{\las \ve{i}\ve{j} \ras}\hat{s}_{\ve{i}\ve{j}}^{x}, \nonumber\\   
\hat{H}_\text{fs}& & =\sum_{\langle \ve{i}\ve{j} \rangle,\sigma}t_{\ve{i}\ve{j}}\xi_{\ve{i}\ve{j}} \hat{s}_{\ve{i}\ve{j}}^{z} \hat{c}_{\ve{i}\sigma}^\dagger \hat{c}^{}_{\ve{j}\sigma}.
  \label{Ham-SM}
\end{eqnarray}
Here $\hat{H}_\text{f} $ corresponds to the half-filled Hubbard model on the $\pi$-flux square lattice. 
$\hat{c}_{\ve{i}\sigma} (\hat{c}_{\ve{i}\sigma}^{\dagger})$ is the fermionic annihilation (creation) operator at site $\ve{i}$ with spin $\sigma = \uparrow, \downarrow$, and $\hat{n}_{\ve{i}\sigma} \equiv \hat{c}_{\ve{i}\sigma}^\dagger \hat{c}_{\ve{i}\sigma}$.
$\hat{H}_\text{s}$ is a ferromagnetic, transverse-field Ising model where the Ising spins---described by the Pauli spin operators
$\hat{s}_{\ve{i},\ve{j}}^{\alpha}(\alpha=x, y, z)$---live on the bonds connecting fermionic sites $\ve{i}$ and $\ve{j}$.
$J$ is the ferromagnetic  nearest-neighbor interaction and $h$ the transverse field.
$\hat{H}_\text{fs}$ accounts for the coupling ($\xi_{\ve{i}\ve{j}}=\pm\xi$) between Dirac fermions and Ising spins.
Our model Hamiltonian $\hat{H}$ has an SU(2) spin symmetry as well as a Z$_2$ symmetry corresponding to invariance under the combined operation of  inversion and $\hat{s}_{\ve{i}\ve{j}}^{z} \to - \hat{s}_{\ve{i}\ve{j}}^{z}$. 
Under inversion $\hat{H}_\text{fs} \to - \hat{H}_\text{fs}$, so that the energy does not depend on the sign of $\xi$ and the two possible Kekul\'e patterns related by $\xi\to-\xi$ are degenerate.  
The Hubbard interaction and the fermion-spin coupling  have the potential to dynamically generate SO(3) AFM order and Z$_2$ KVBS order (ferromagnetic order of the Ising spins), respectively.

\begin{figure}
\centering
\centerline{\includegraphics[width=0.45\textwidth]{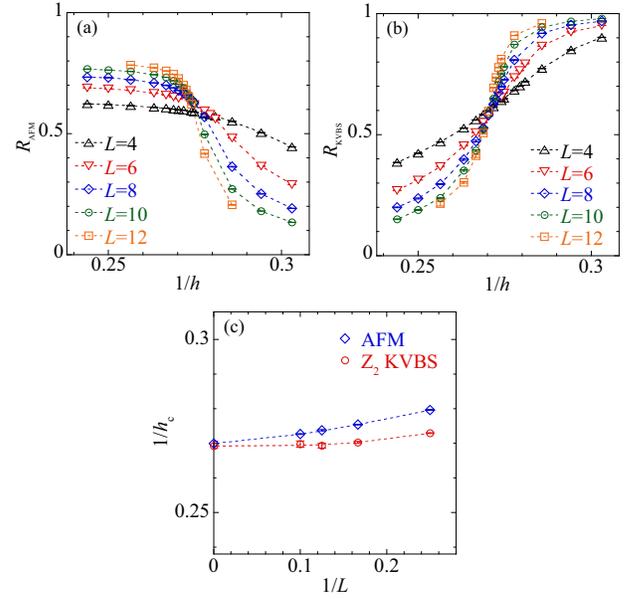}}
\caption{\label{fig:Fig-2-SM} Correlation ratios for (a) AFM and (b) $\mathrm{Z}_2$ KVBS states at $U=7$. 
(c) Extrapolation of crossing points of correlation ratios for $L$ and $L+2$ gives the critical value $\left(h_c^{\text{AFM}}\right)^{-1}\approx \left(h_c^{\text{KVBS}}\right)^{-1}\approx 0.270$.}
\end{figure}

We used the ALF (Algorithms for Lattice Fermions) implementation \cite{ALF_v1} of the well-established finite-temperature auxiliary-field QMC method~\cite{Blankenbecler81,Assaad08_rev}. 
Simulations of our model are free of the negative sign problem.  To see this, one can first  carry out a partial  particle-hole canonical transformation:
$\hat{c}^{}_{\ve{i},\uparrow} \rightarrow e^{i \ve{Q}\cdot \ve{i}}\hat{c}^{\dagger}_{\ve{i},\uparrow}$ and $\hat{c}^{}_{\ve{i},\downarrow} \rightarrow \hat{c}^{}_{\ve{i},\downarrow}$   with $\ve{Q} = (\pi,\pi)/a$. This changes the sign of the Hubbard interaction from positive to negative. One can the use time reversal symmetry to show that the eigenvalues of the Fermion determinant come in complex conjugate pairs \cite{Wu04}.
We simulated lattices with $L\times L$ unit cells (each containing four Dirac fermions and four Ising spins) and periodic boundary conditions.
Henceforth, we use $t=1$ as the energy unit, set $J=-1$, $\xi=0.5$, and $U=7$.
All the data are for the Trotter discretization $\Delta\tau=0.1$.
In the considered parameter range, an inverse temperature $\beta=30$ was sufficient to obtain results representative of the ground state.

\begin{figure}
\centering
\vspace{0cm}
\centerline{\includegraphics[width=0.45\textwidth]{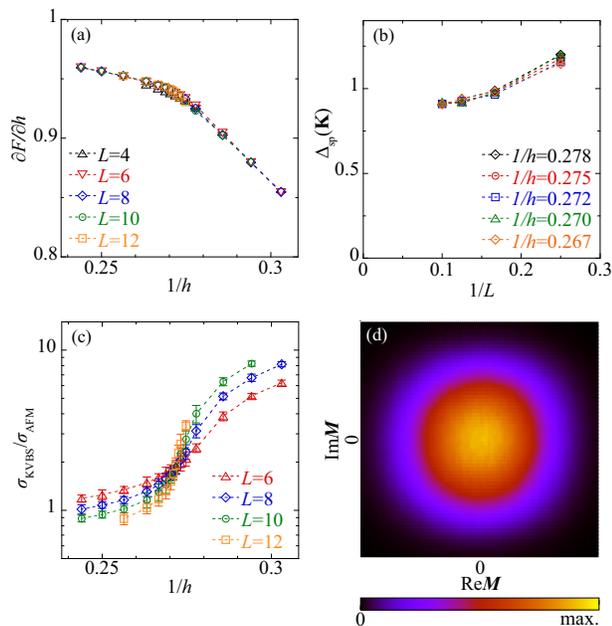}}
\caption{(a) Free-energy derivative and (b) single-particle gap at the Dirac point $\ve{q}= \ve{K}$. 
(c) Ratio of the standard deviations of the AFM and Z$_2$ KVBS order parameters.
(d) Joint probability distribution of the two order parameters at the critical point for $L=8$. Here, $U=7$ and $\beta=30$.}  
\label{fig:Fig-3-SM}
\end{figure}

The dashed line in Fig. 1 (a) of the main text indicates the scan of the
phase diagram obtained by tuning $h$ at a fixed $U=7$.  Along this line we
observe two insulating phases corresponding to the SO(3) AFM and the Z$_2$
KVBS.

To map out the phase diagram we compute equal-time correlation
functions of fermion spin
$\hat{\vec{ S}}_{\ve{i}}=\sum_{\sigma\sigma'}
\hat{c}_{\ve{i}\sigma}^{\dagger}\vec{\sigma}_{\sigma\sigma'}
\hat{c}^{}_{\ve{i}\sigma'}$, fermion bond
$\hat{B}_{\ve{i}\ve{j}}=\sum_\sigma t_{\ve{i}\ve{j}}
(\hat{c}_{\ve{i}\sigma}^\dagger
\hat{c}^{}_{\ve{j}\sigma}+\hat{c}_{\ve{j}\sigma}^\dagger
\hat{c}^{}_{\ve{i}\sigma})$, and Ising spin $\hat{s}_{\ve{i}\ve{j}}^{z}$.
Due to the larger unit cell, these correlation functions are $4\times 4$
matrices of the form
$C^O_{\ve{R}\gamma,\ve{R}'\delta}=\langle (\hat{{O}}_{\ve{R}\gamma}-\langle
\hat{{O}}_{\ve{R}\gamma} \rangle)\cdot(\hat{{O}}_{\ve{R}'\delta}-\langle
\hat{{O}}_{\ve{R}'\delta} \rangle)\rangle$
($\hat{O}=\hat{\vec{S}},\hat{B},\hat{s}$) where $\ve{R}, \ve{R}'$ labels the
unit cell and $\gamma,\delta$ the orbitals.  After diagonalizing the
corresponding structure factors
\begin{equation}
C^O_{\gamma  \delta}(\vec{q})=\frac{1}{L^2}\sum_{\ve{R}\ve{R}'}C^O_{\ve{R}\gamma,\ve{R}'\delta}e^{i {\vec q}\cdot (\ve{R}-\ve{R}')},
\label{eq:SF-SM}
\end{equation}
we calculated the renormalisation-group invariant correlation ratio~\cite{Binder1981,Pujari16}
\begin{equation}
R_{O}=1-\frac{\lambda_1({\ve q}_0+\delta {\ve q})}{\lambda_1({\ve q}_0)}
\label{eq:CR-SM}
\end{equation}
using the largest eigenvalue $\lambda_1({\ve q})$; ${\ve q}_0$ is the ordering wave vector, ${\ve q}_0 + \delta {\ve q}$ a neighboring wave vector.  
By definition, $R_O\to 1$ for $L\to\infty$ in the corresponding ordered state, whereas $R_O\to 0$ in the disordered state.  
At the critical point, $R_O$ is scale-invariant for sufficiently large $L$ so that results for different system sizes cross.
Figures~\ref{fig:Fig-2-SM} (a) and (b) show the results.
Based on the assumption of a dynamical critical exponent $z=1$~\cite{Herbut09}  we set $L=\beta$  to carry out the finite-size scaling. 
The onset of the AFM is detected from the crossing of $R_\text{AFM}\equiv R_{S}$ [Fig.~\ref{fig:Fig-2-SM}(a)], whereas the onset of Kekul\'e order can be detected either from $R_\text{KVBS}\equiv R_{s}$ [Fig.~\ref{fig:Fig-2-SM}(b)] or from $R_B$.  
The finite-size scaling of the crossing points yields a single critical point of $h_c^{-1}\approx0.270$ shown in Fig.~\ref{fig:Fig-2-SM} (c), pointing to a direct phase transition between AFM and Z$_2$ KVBS.

Figure~\ref{fig:Fig-3-SM} (a) plots the free-energy derivative
\begin{equation}
\frac{\partial F}{\partial h}=  \frac{1}{4L^2}  \left< \sum_{\langle \ve{i}\ve{j} \rangle}\hat{s}_{\ve{i}\ve{j}}^{x}  \right>. 
\label{eq:DF-SM}
\end{equation}
The absence of a discontinuity in this quantity favors  a continuous transition.  

For the derivation of a low-energy effective field theory is important to confirm that the single particle gap remains finite across the  transition.
We measured the imaginary-time displaced Green's function $G({\ve q},
\tau)$ and obtained the single-particle gap $\Delta_{\text{sp}}({\ve q})$ from
\begin{equation}
G({\ve q}, \tau) \propto \exp(-\tau\Delta_{\text{sp}}({\ve q}) )
\label{eq:SG-SM}
\end{equation}
at large imaginary time $\tau$.
We found that the single-particle gap at the Dirac point, shown in Fig.~\ref{fig:Fig-3-SM}(b), remains clearly nonzero.

To verify whether the critical point has an emergent SO(4) symmetry, we measured the standard deviations $\sigma_O=\sqrt{\las \hat{O}^2\ras-\las \hat{O}\ras^2}$ of the AFM ($\sigma_{\text{AFM}}\equiv\sigma_{S}$) and Z$_2$ KVBS ($\sigma_{\text{KVBS}}\equiv\sigma_{s})$ order parameters.
These quantities are in general independent but become locked together and
can be combined into a four-component order parameter if an SO(4)
symmetry---unifying the two order parameters---emerges at the critical point.
In this case, the ratio $\sigma_{\rm KVBS}/\sigma_{\rm AFM}$ will become universal at the critical point.
This is confirmed by the result in Fig.~\ref{fig:Fig-3-SM}(c).
Moreover, the emergent  SO(4) symmetry can also be confirmed in the joint probability distribution of the two order parameters determined from QMC snapshots,
\begin{equation}
\boldsymbol{M}=\sigma_{\text{AFM}}e^{i0}+\sigma_{\text{KVBS}}e^{\frac{i\pi}{2}}.
\label{eq:HG-SM}
\end{equation}
If the SO(4) symmetry emerges at the critical point, the quantity
$\boldsymbol{M}$ (after normalization of each order parameter) should reveal a circular distribution, as confirmed  at the critical point
by Fig.~\ref{fig:Fig-3-SM}(d).

\section{Spin dynamics around the Z$_2$ domain wall with varying $1/h$}
\label{QMC-Cs.sec}

\begin{figure}
\centering
\centerline{\includegraphics[width=0.45\textwidth]{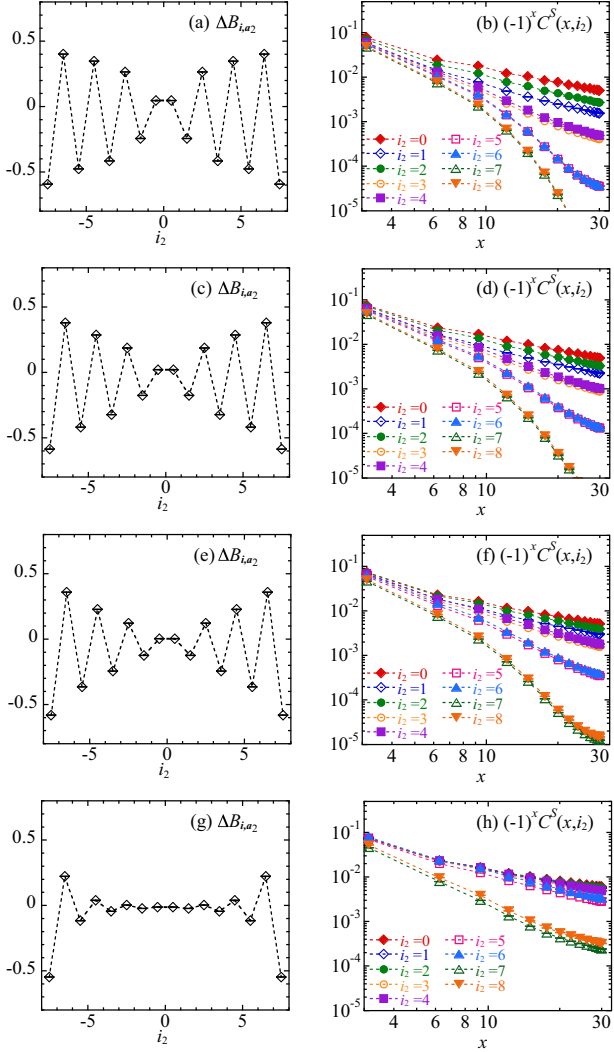}}
\caption{\label{fig:Fig-5-SM}
Real-space bond energy change $\Delta \hat{B}_{\ve{i},\ve{a}_2}$ at $\ve{i}=(i_1=0, i_2)$ (left panel) and real-space spin-spin correlation functions along the domain wall of $\mathrm{Z}_2$ KVBS $C^S(x, i_2)$ (right panel)  for $L_1=30$ and $L_2=17$. Here $x=L_1  {\rm sin}(\pi i_1/L_1)$ is the conformal distance. Parameters used are $1/h=0.333$ [(a) and (b) in the $\mathrm{Z}_2$ KVBS phase],  $1/h=0.299$ [(c) and (d) in the  $\mathrm{Z}_2$ KVBS phase], $1/h=0.275$ [(e) and (f) in the  $\mathrm{Z}_2$ KVBS phase close to the critical point], and $1/h=0.033$ [(g) and (h) in the AFM phase]. Here, $U=7$ and $\beta=30$.
}
\end{figure}
In the main text  we  set  $1/h=0.275$ that places us in the  $\mathrm{Z}_2$ KVBS phase close to the critical point.  
Here  we vary $1/h$  across  the transition  and investigate the fate of the domain wall.   
In Fig.~\ref{fig:Fig-5-SM}   we  map out the profile of the domain wall by considering  $\Delta \hat{B}_{\ve{i},\ve{a}_2}$ across the domain wall (left panels)  the spin correlations  along the domain wall (right panels).  These quantities are defined in the main text.

The profile of the domain wall  in the KVBS  phase is expected to be inversely proportional to the stiffness.  Indeed the QMC results shown in Figs.~\ref{fig:Fig-5-SM} (a)-(f) support  that upon moving away from the critical point \textit{deep}  into  the  $\mathrm{Z}_2$ KVBS phase the  domain wall becomes more pronounced.   Accordingly the width around the domain wall   where  one observes  $1/r$ decay of the spin-spin correlations reduces.      

In the AFM phase  there is no scale  that confines the  width of the domain wall other than the width of the lattice  $L_2$.       We equally expect the spin-spin correlations to show long range order.  In  Fig.~\ref{fig:Fig-5-SM} (g)-(h)    we  consider a very large value of $h=30$. As apparent from Fig.~\ref{fig:Fig-5-SM} (g)    the profile of the domain wall is very flat   around the center of the  cylinder.    The spin correlations show a  slight upturn  but remarkably do  not provide clear evidence of   long range order.  We can understand this in the following way.  
The AFM state originates from the binding of spinons  and this will occur on a given length scale.  If   $L_2$  is comparable to this length scale, then long range order will be hard to detect.  We hence conjecture that as $L_2$ grows  spin-spin correlations will develop clear signs 
of ordering.

\section{One dimensional Hubbard model at half filling}
\label{1DQMC.sec}

In this section, we present  QMC results of the one-dimensional repulsive
Hubbard model at half filling. The Hamiltonian is
\begin{eqnarray}
\hat{H}=t\sum_{i,\sigma}\hat{c}_{i\sigma}^\dagger \hat{c}^{}_{i+1\sigma}  + H.c. +U\sum_{i}(\hat{n}_{i\uparrow}-\oh)(\hat{n}_{i\downarrow}-\oh), \nonumber \\
  \label{Ham1D-SM}
\end{eqnarray}
where $t$ is the nearest-neighbor hopping amplitude and $U$ is the Hubbard repulsion.
$\hat{c}_{i\sigma} (\hat{c}_{i\sigma}^{\dagger})$ is the fermionic annihilation (creation) operator at site $i$ with spin $\sigma = \uparrow, \downarrow$, and $\hat{n}_{i\sigma} \equiv \hat{c}_{i\sigma}^\dagger \hat{c}_{i\sigma}$.
For the numerical simulations we used the ALF (Algorithms for Lattice Fermions) implementation~\cite{ALF_v1} of the well-established auxiliary-field quantum Monte Carlo (AFQMC) method~\cite{Blankenbecler81,Assaad08_rev}.
We carried out ground-state simulations with the projective AFQMC algorithm, which is based on the equation
\begin{eqnarray}
 \langle \hat O \rangle_{\hat{H}} =\lim_{\Theta \to \infty} \frac{\langle \psi_T | e^{-\frac{\Theta}{2} \hat H }  \hat {O} e^{- \frac{\Theta}{2} \hat H}  | \psi_T  \rangle }
      {   \langle \psi_T |   e^{- {\Theta} \hat H}  | \psi_T  \rangle     },
        \label{CF-1D}
\end{eqnarray}
where $\Theta$ is a projection parameter. The trial wave function $ | \psi_T \rangle $  is chosen to correspond to the ground state of the noninteracting Hamiltonian. 
We simulated lattice sizes ranging from $L=14$ to $510$  with periodic boundary conditions.
Henceforth, we use $t=1$ as the energy unit and set $U=4$.
All the data were obtained for $\Theta=30$ (and Trotter discretization
$\Delta\tau=0.1$),  sufficient to
obtain results representative of the ground state for the parameters considered.

\begin{figure}
\centering
\centerline{\includegraphics[width=0.45\textwidth]{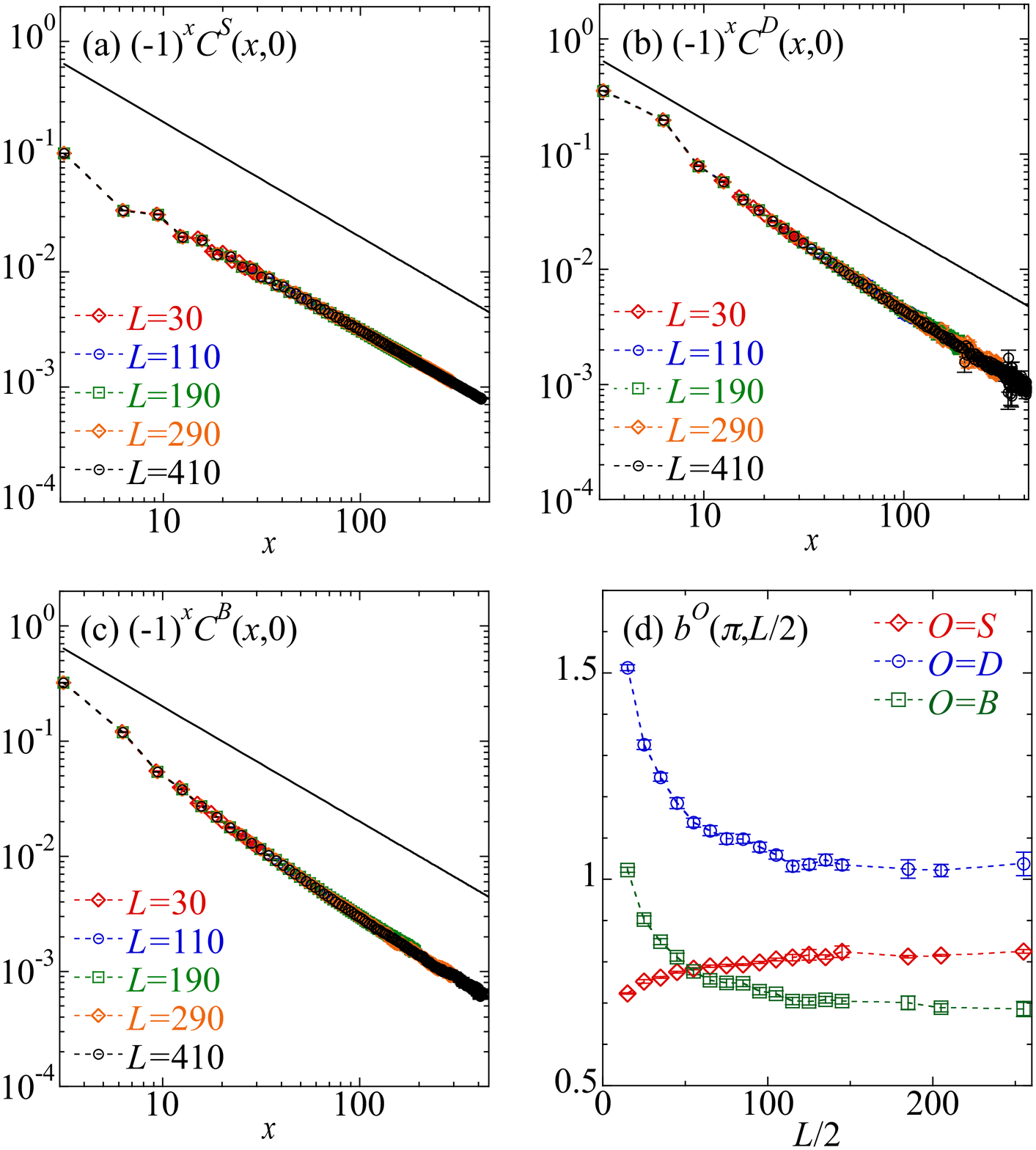}
}
\caption{\label{fig:CF-1D-SM}
Real-space correlation functions of the one-dimensional repulsive Hubbard
model at half filling for (a) spin, (b) dimer, and (c) bond. Here $x=L{\rm
  sin}(\pi r/L)$ is the conformal distance. Solid line correspond to $(-1)^x x^{-1}$.
(d) $b^S(q, L/2)$, $b^D(q, L/2)$, and $b^B(q, L/2)$ at $q=\pi$ (see text).
}
\end{figure}

We measured real-space correlation functions of spin $\hat{{\vec S}}_{i}=\sum_{\sigma\sigma'} \hat{c}_{i \sigma}^{\dagger} \vec{\sigma}_{\sigma\sigma'} \hat{c}^{}_{i \sigma'}$, dimer $\hat{{\vec D}}_{i}=\hat{{\vec S}}_{i}\cdot\hat{{\vec S}}_{i+1}$, and bond $\hat{B}_{i}=\sum_\sigma (\hat{c}_{i \sigma}^\dagger \hat{c}^{}_{i+1\sigma}+\hat{c}_{i+1\sigma}^\dagger \hat{c}^{}_{i\sigma})$,
\begin{eqnarray}
C^S(r)&&=\langle \hat{\vec S}_{r}\cdot\hat{\vec S}_{0}\rangle,\\ 
C^D(r)&&=\langle (\hat{\vec D}_{r}-\langle \hat{\vec D}_{r} \rangle)\cdot(\hat{\vec D}_{0}-\langle \hat{\vec D}_{0} \rangle)\rangle,\\ 
C^B(r)&&=\langle (\hat{B}_{r}-\langle \hat{B}_{r} \rangle)\cdot(\hat{B}_{0}-\langle \hat{B}_{0} \rangle)\rangle.
  \label{CF-1D}
\end{eqnarray}
The ground state of the SU(2) symmetric isotropic spin-$1/2$ Heisenberg chain is conformally invariant.
Spin correlation functions scale as $\sim (-1)^r (\ln{r})^{1/2}r^{-1}$, with multiplicative logarithmic corrections~\cite{Affleck_1989,Singh89,Giamarchi89}.
Owing to emergent SO(4) symmetry  \cite{Affleck85,Affleck87}, dimer
correlation functions exhibit the same exponent but different logarithmic
corrections; they decay as  $ \sim(-1)^r (\ln{r})^{-3/2}r^{-1}$~\cite{Giamarchi89}.
Here we consider the  half-filled Hubbard  Hamiltonian that maps onto the Heisenberg model.  

Figures~\ref{fig:CF-1D-SM}(a) and (b) show QMC results for spin and dimer correlation functions, which indicate that the correlation functions both yield  consistent power-law decay at large conformal distances, $x=L{\rm sin}(\pi r/L)$.
The bond correlations have the same symmetry properties as the dimer correlations so that we expect the same power-law decay. 
This can be confirmed by the QMC results shown in Fig.~\ref{fig:CF-1D-SM} (c).
The exponent of the power-law decay  can be detected using the  quantity~\cite{Assaad91}
\begin{eqnarray}
b^O(q, L/2)&&\equiv C^O(q)\nonumber\\ 
&&-C^O(q+2\pi/L)-C^O(q-2\pi/L),
  \label{SF}
\end{eqnarray}
where  $C^O(q)$  corresponds to the static structure factor of the local observable $O$.
One can show that the real space correlations at distance  $C^O(L/2)
=\frac{1}{4L}\sum_{q} e^{iqL/2}b^O(q, L/2)$.  This formula singles out  the
wave numbers  where the static structure factor shows non-analytical
behavior.  In our case, this is the case at $q=\pi$ so that 
\begin{equation} 
C^O(L/2)  \propto \frac{1}{L}  e^{i \pi L/2}b^O(q=\pi, L/2).
\end{equation}
Assuming that
$C^O(L/2) \propto \alpha e^{i \pi L/2} (L/2)^{-K} (\ln(L/2))^{\gamma} $  then 
\begin{equation}
\ln(b^O(\pi, L/2)) \propto \alpha'+(-K + 1) \ln(L/2)+\gamma \ln(\ln(L/2)).
  \label{SF}
\end{equation}
Figure~\ref{fig:CF-1D-SM}(d) plots this quantity for spin, dimer, and bond  correlations as a function of $L/2$.
All  three quantities are nearly constant for large values of $L/2$,  thus confirming  $K=1$. 
Differences in  $b^D(\pi, L/2)$  and  $b^S(\pi, L/2)$ arise  from the third term on the right-hand side of Eq.~(\ref{SF}),  reflecting the  different logarithmic corrections.
Since bond correlations have the same symmetry properties as the dimer correlations, the logarithmic corrections are expected to be consistent. 

\end{document}